\documentclass{PoS}
\usepackage{wrapfig}

\title{Science Highlights from VERITAS}

\ShortTitle{VERITAS Science Highlights}

\author{\speaker{D. Staszak} for the VERITAS Collaboration\thanks{See http://veritas.sao.arizona.edu/conferences/authors?icrc2015 for a full list of affiliations} :
A.~U.~Abeysekara,
S.~Archambault,
A.~Archer,
A.~Barnacka,
W.~Benbow,
R.~Bird,
J.~Biteau,
M.~Buchovecky,
J.~H.~Buckley,
V.~Bugaev,
K.~Byrum,
J.~V~Cardenzana,
M.~Cerruti,
X.~Chen,
J.~L.~Christiansen,
L.~Ciupik,
E.~Collins-Hughes,
M.~P.~Connolly,
P.~Coppi,
W.~Cui,
V.~V.~Dwarkadas,
J.~D.~Eisch,
M.~Errando,
A.~Falcone,
Q.~Feng,
M.~Fernandez~Alonso,
J.~P.~Finley,
H.~Fleischhack,
A.~Flinders,
P.~Fortin,
L.~Fortson,
A.~Furniss,
G.~H.~Gillanders,
S.~Griffin,
S.~T.~Griffiths,
G.~Gyuk,
M.~H{\"u}tten,
D.~Hanna,
J.~Holder,
T.~B.~Humensky,
C.~A.~Johnson,
P.~Kaaret,
P.~Kar,
M.~Kertzman,
Y.~Khassen,
D.~Kieda,
M.~Krause,
F.~Krennrich,
S.~Kumar,
M.~J.~Lang,
G.~Maier,
S.~McArthur,
A.~McCann,
K.~Meagher,
P.~Moriarty,
R.~Mukherjee,
T.~Nguyen,
D.~Nieto,
A.~O'Faol\'{a}in de Bhr\'{o}ithe,
R.~A.~Ong,
A.~N.~Otte,
D.~Pandel,
N.~Park,
V.~Pelassa,
J.~S.~Perkins,
A.~Petrashyk,
M.~Pohl,
A.~Popkow,
E.~Pueschel,
J.~Quinn,
K.~Ragan,
P.~T.~Reynolds,
G.~T.~Richards,
E.~Roache,
A.~C.~Rovero,
M.~Santander,
S.~Schlenstedt,
G.~H.~Sembroski,
K.~Shahinyan,
A.~W.~Smith,
I.~Telezhinsky,
J.~V.~Tucci,
J.~Tyler,
V.~V.~Vassiliev,
S.~Vincent,
S.~P.~Wakely,
O.~M.~Weiner,
A.~Weinstein,
A.~Wilhelm,
D.~A.~Williams,
B.~Zitzer
	\\
        McGill University\\
        E-mail: \email{staszak@physics.mcgill.ca}}

\abstract{

The Very Energetic Radiation Imaging Telescope Array System (VERITAS) is a ground-based array located at the Fred Lawrence Whipple Observatory in southern Arizona and is one of the world's most sensitive gamma-ray instruments at energies of 85 GeV to $>$30 TeV.
VERITAS has a wide scientific reach that includes the study of extragalactic and Galactic objects as well as the search for astrophysical signatures of dark matter and the measurement of cosmic rays.
In this paper, we will summarize the current status of the VERITAS observatory and present some of the scientific highlights from the last two years, focusing in particular on those results shown at the 2015 ICRC in The Hague, Netherlands.
}

\FullConference{The 34th International Cosmic Ray Conference,\\
		30 July- 6 August, 2015\\
		The Hague, The Netherlands}

\begin{document}

\section{Introduction}

The Very Energetic Radiation Imaging Telescope Array System (VERITAS) is a ground-based array located at the Fred Lawrence Whipple Observatory in southern Arizona and is one of the world's most sensitive gamma-ray instruments at energies of 85 GeV to $>$30 TeV.
VERITAS has a wide scientific reach that includes the study of extragalactic and Galactic objects as well as the search for astrophysical signatures of dark matter and the measurement of cosmic rays.
In this paper, we will summarize the current status of the VERITAS observatory and present some of the scientific highlights from the last two years, focusing in particular on those results shown at the 2015 ICRC in The Hague, Netherlands.

\begin{center}
\begin{figure}
\begin{center}
\includegraphics[width = 5.2in]{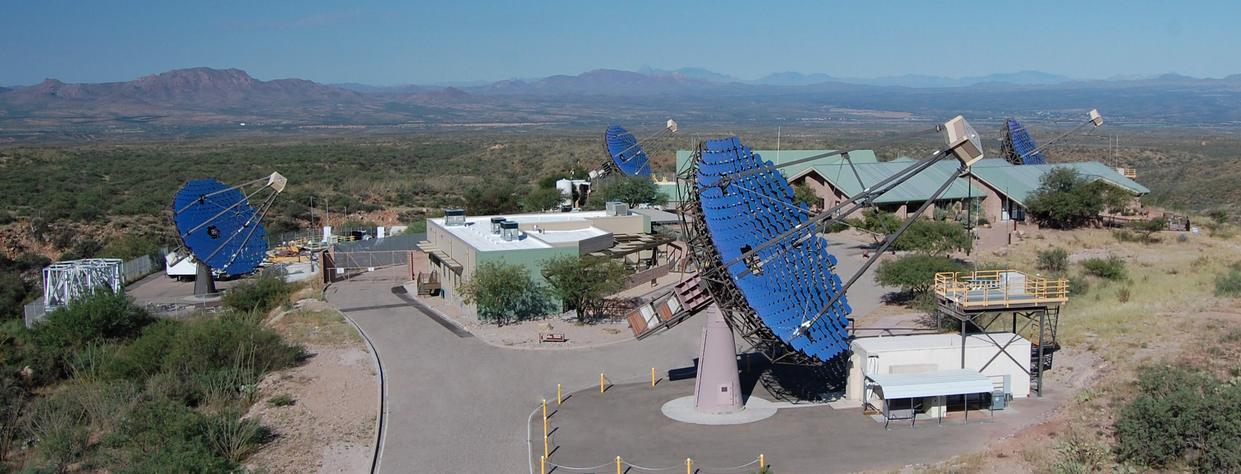}
\end{center}
\includegraphics[width = 3in]{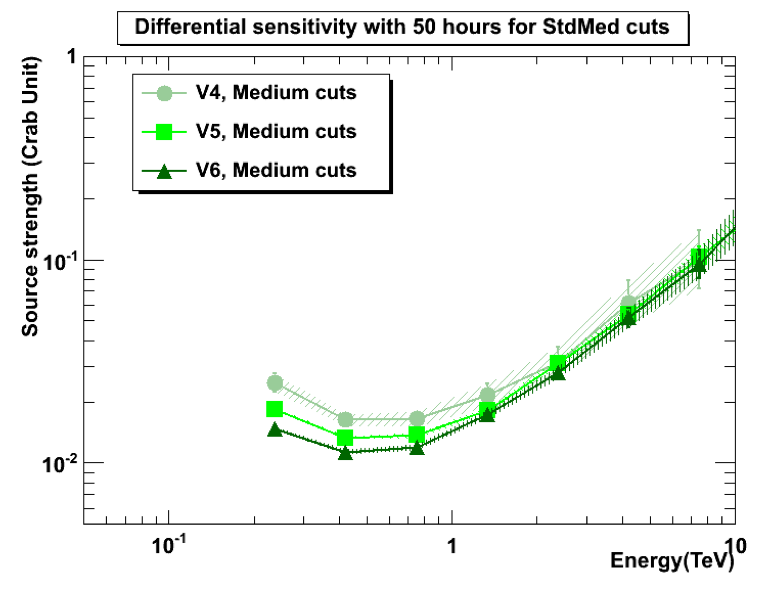}
\includegraphics[width = 3in,height=2.5in]{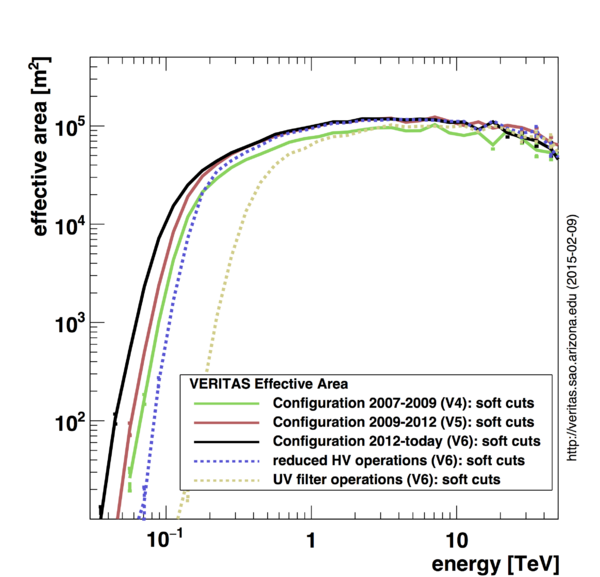}
\caption{
At the top the VERITAS array is shown in its present configuration.
Shown in the bottom left panel is the VERITAS differential sensitivity using similar cuts in all three hardware configurations: V4 - prior to T1 move, V5 - after T1 move but before PMT upgrade, V6 - after PMT upgrade.  
Shown in the bottom right panel are VERITAS effective areas for these hardware configurations as well as for the two bright moonlight observing configurations.}
\label{sensPlot}
\end{figure}
\end{center}

\begin{sloppypar}
The VERITAS array was completed in 2007 and consists of four imaging atmospheric Cherenkov telescopes (IACTs) with typical baselines of $\sim$100 m between telescopes\cite{holder}.
Each telescope comprises a 12 m reflector and an imaging camera that is instrumented with 499 PMTs, creating a 3.5$^{\circ}$ field of view on the sky.
Since inauguration, the array has undergone two major hardware upgrades. 
The first, in 2009, relocated one of the telescopes to better symmetrize the array\cite{T1move}.
This upgrade, when combined with a newly developed mirror alignment technique\cite{alignment}, improved the angular resolution of the instrument and brought the sensitivity for detecting a 1$\%$ Crab Nebula source down from 50 hours of observation to 30 hours.
The second, over 2011-2012, replaced the L2 trigger system\cite{L2paper} and instrumented each of the cameras with new, high efficiency PMTs\cite{newPMTs}.
The new PMTs collect significantly more light and have impacted the VERITAS science programs where low energy sensitivity is critical. 
In particular, we have now detected several new soft spectrum AGN that would not have been possible under the original array configuration.
Figure \ref{sensPlot} shows the array in its current configuration and summarizes the overall sensitivity of these three hardware configurations\cite{nahee}.
In the current configuration, VERITAS can detect a 1$\%$ Crab Nebula source in $\sim$25 hours and has an angular resolution of 0.1$^{\circ}$ at 1 TeV (68$\%$ containment), an energy resolution of $15-25\%$, and a pointing accuracy of less than 50 arcsec.
\end{sloppypar}

VERITAS typically collects $\sim$1400 hours of data per season, with two months of monsoons in the summer where observation is impossible.
Of this total, low moonlight observations, characterized by the moon in the sky but with moon illumination less than $30\%$, accounts for $\sim$170 hours.
Additionally, VERITAS has actively pursued observations into bright moonlight conditions ($\sim$300 hours), where the moon is illuminated to $>$50$\%$\cite{MOON}.
We collect data under bright moonlight conditions using two non-standard techniques, one with reduced high voltage PMT settings and the other with UV filters placed over the PMTs.
The utility of aggressively adding observation time under these conditions was demonstrated by the detection of a flare state in a BL Lac object, 1ES 1727+502\cite{1727}, under bright moonlight conditions.
This represents the first such publication using bright moonlight data by VERITAS.

VERITAS is now a mature scientific instrument that has completed 8 years of successful data collection with minimal hardware problems.
The VERITAS source catalog now lists 54 objects from at least eight different source classes.
The scientific focus of VERITAS has evolved from source discovery to collecting deep exposures on known objects that yield the most scientific bounty.
Last year the collaboration developed a {\it Long Term Plan} to map out the scientific priorities for the next five years.
The plan reserves $70\%$ of our average yearly observational time for our core science plan, which includes a mix of sources that offer the most scientific impact within each of the main VERITAS science programs.
The remaining time is opened up at an annual time-allocation competition.
Additionally, up to $15\%$ of the total VERITAS observation time is accessible, with funding, to external scientists through the {\it Fermi}-VERITAS-GI program\footnote{http://fermi.gsfc.nasa.gov/ssc/proposals/veritas.html}.

\section{Extragalactic Science}

Extragalactic sources account for $\sim$40$\%$ of VERITAS observational time and $\sim$65$\%$ of all VERITAS detections.
All of these detections are BL Lac objects with three exceptions, the starburst galaxy M82 and the two radio galaxies M87 and NGC 1275.	
VERITAS now has exposures on over 100 northern blazars and pursues an active multi-wavelength monitoring program to detect flares in known and candidate objects.
BL Lacs are known to exhibit flux variation on the scale of minutes to hours to years, and catching an object in a high state may be a means to discover VHE (Very High Energy, VHE:$>$100 GeV) emission where otherwise its quiescent state may be too dim.
Further, besides a means of discovering new sources, flares provide a way to accumulate large gamma-ray statistics on a given object in a short period of time.
This can be used to push spectral measurements to higher energies where many cosmological quantities can be probed.

Multi-wavelength (MWL) data is critical to blazar science during both flaring and quiescent states, and VERITAS has many ongoing MWL programs and partnerships.
Broadband modelling of the spectral energy distribution (SED) provides a means to study the relativistic particle populations within the source jets, and to distinguish between different gamma-ray emission models.
We ensure all VERITAS blazar detections have MWL data.
To date, most VERITAS SEDs are found to be compatible with the synchrotron self-compton (SSC) emission models, though a few slightly favor other mechanisms. 

The scientific focus of blazar observations at VERITAS has historically been a mix of source discovery and deep exposures on known objects.
To maximize the scientific output of the blazar program, VERITAS now focuses the majority of its observational time on deep exposures of a select group of sources.
This program is discussed in more detail elsewhere in these proceedings\cite{wystan}, here we will focus on a few recent highlights and ongoing studies. 

The last six months have been particularly active for northern gamma-ray blazars, resulting in several new VERITAS detections of potentially interesting objects: PKS 1441+25, RGB J2243+203, and S3 1227+25.  
PKS 1441+25 is the second flat spectrum radio quasar (FSRQ) detected by VERITAS and the fifth at VHE energies.  
VERITAS observed this object for $\sim$15 hours over the course of a week starting the night of April 15th, 2015, and found it to exhibit a steady $\sim$5$\%$ Crab Nebula flux above 80 GeV.
VERITAS observations were triggered by a detection the previous night from MAGIC, which were themselves triggered by monitoring and alert emails from the $Fermi$-LAT team\cite{PKSatel}.
The redshift for this object is particularly distant, z=0.939, making preliminary Extragalactic Background Light (EBL) constraints for this flare competitive.
RGB J2243+203 was observed for 280 minutes between Dec. 21st and Dec. 24th, 2014, resulting in a cumulative significance of 5.6$\sigma$\cite{2243atel}\cite{udara}. 
The redshift of this object is unknown (photometrically estimated at z$>0.39$) but it was identified by $Fermi$-LAT as a good candidate for VHE observation in their hard source catalog, 1FHL J2244.0+2020\cite{1FHL}.
VERITAS has an active program to monitor the $Fermi$-LAT data of known and potential VHE emitters\cite{monitorLAT1}\cite{monitorLAT2} and this source was targeted for observations after it exhibited a flux increase in the 1-100 GeV band.
Finally, S3 1227+25 (z=0.135) is an LBL/IBL that was detected at $\sim10\sigma$ in 6 hours of VERITAS exposure\cite{S3atel}.
Unlike other recent flare detections, this source is rather soft and was not assumed to be a good VHE candidate ($\Gamma_{1FHL} \sim 3.3$).
VERITAS observations were triggered by an email alert from the $Fermi$-LAT team.

One of the biggest blazar science stories over the last couple of years has been the detection of VHE gamma-ray sources that are more and more distant.
VERITAS has been at the center of this story.
Most VHE-detected blazars are relatively close (with redshifts of $z<0.3$), but VERITAS has been collecting a sizable list of more distant objects: 3C 66A (0.33$<$z$<$0.41), PKS 1222+16 (z=0.432), PG 1553+113 (0.43$<$z$<$0.58), PKS 1424+240 (z$>$0.60), as well as the new detection of PKS 1441+25 (0.939).
Distant blazars are cosmologically interesting because VHE gamma rays are expected to interact with intervening low-energy extragalactic background light (EBL) via pair creation, $\gamma_{EBL}+\gamma_{VHE}\rightarrow e^{+}+e^{-}$, effectively creating a gamma-ray "horizon".
The probability of this interaction is characterized by an exponential attenuation factor, $e^{-\tau}$, where $\tau$ is the optical depth and depends on the object's redshift and the gamma-ray energy.
Of particular utility in these studies is the fact that gamma rays below $\sim$100 GeV are not expected to be significantly absorbed, so a comparison of energy spectra at $Fermi$-LAT and VERITAS energies can be used to extract information about the intrinsic and absorbed gamma-ray spectra, respectively.

\begin{figure}
\begin{center}
\includegraphics[width = 4in]{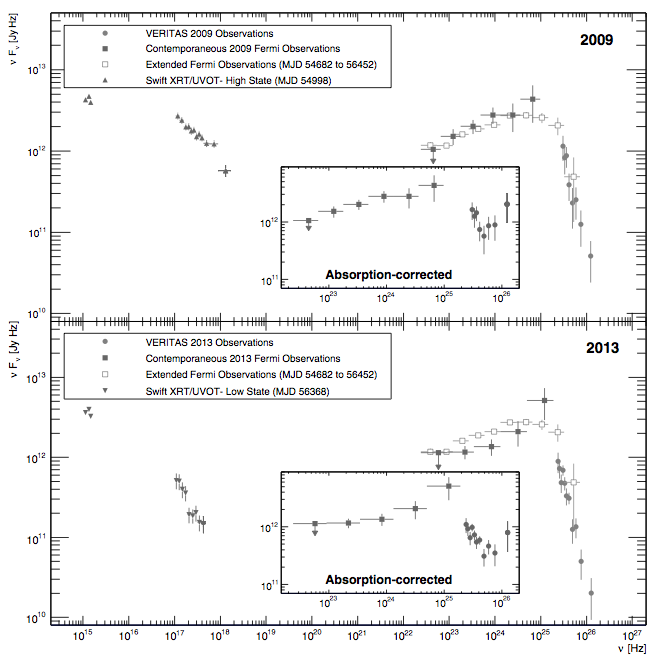}
\caption{Broadband PKS 1424+240 SEDs for VERITAS data collected in 2009 and 2013, corresponding to relatively high and low states, respectively. 
Swift and $Fermi$-LAT data are additionally shown, where both contemporaneous and the full set of $Fermi$ data is also presented. 
Insets show the deabsorbed spectral points for both datasets.
}
\label{1424Plot}
\end{center}
\end{figure}

PKS 1424+240 is a HBL discovered by VERITAS in 2009 that was recently found to be very distant, with $z>0.6$\cite{amy1424}.
The early 2009 VERITAS dataset was complemented by a deep 2013 dataset and both were analyzed over the $\sim1-750$ GeV energy range\cite{1424}.
Measured and deabsorbed spectra for these datasets are shown in Fig. \ref{1424Plot} using the EBL model from Gilmore\cite{gilmore}.
Unfortunately, the flux in 2013 was less than in 2009, so while the detection of the source is improved with the factor of $\sim2.3$ in exposure(14.4$\sigma$ vs. 8.5$\sigma$), we did not extend the energy for which there was a spectral measurement.
In both datasets a marginal hardening in the spectrum at higher energies is seen after correcting for EBL absorption.
A high energy spectral hardening can be evidence of an overestimation of the EBL density or some more exotic physical mechanism, such as internal photon-photon absorption or gamma-ray production from secondary cascades caused by cosmic-ray line-of-sight interactions.

VERITAS blazar measurements can also be used to indirectly probe the magnetic fields that fill the voids between the large-scale structures of the universe.
These fields, known collectively as the intergalactic magnetic field (IGMF), are believed to exist but are difficult to measure directly, leaving their magnitude and origin as a critical unresolved problem.
As described above, gamma rays of order of hundreds of GeV are expected to interact with the EBL, creating an electron-positron pair whose trajectory is sensitive to the IGMF.
These electrons can then up-scatter cosmic microwave background (CMB) photons to create a secondary cascade of lower energy gamma rays.
Cascade gamma rays would exhibit temporal and spatial properties imprinted by the strength of the IGMF, such as a time-delay and an angular broadening of the gamma-ray signal beyond the instrument PSF.

\begin{figure}
\begin{center}
\includegraphics[width = 1.96in]{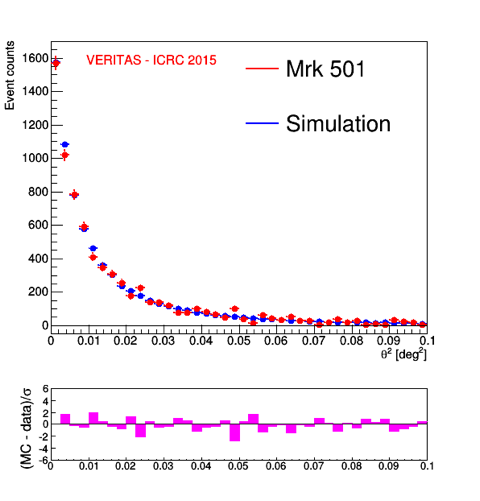}
\includegraphics[width = 1.96in]{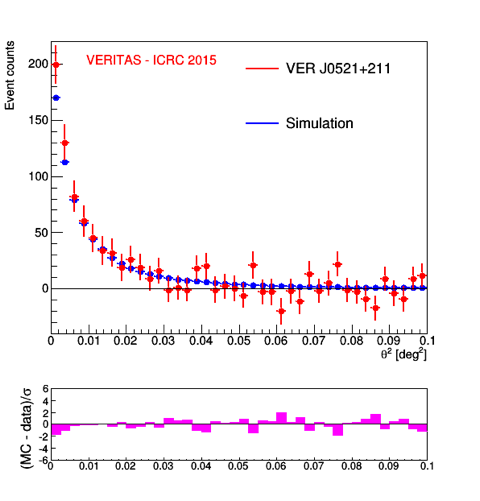}
\includegraphics[width = 1.96in]{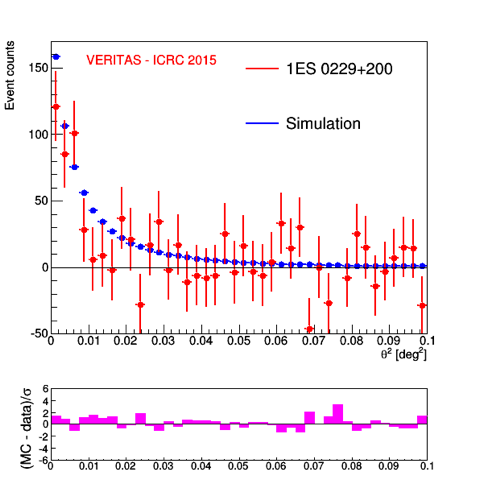}
\caption{
The angular distribution ($\theta^{2}$) for three of the VERITAS blazars considered in \cite{elisa}: Mrk 501, VER J0521$+$211, and 1ES 0229$+$200.
Shown in the top panels are the data in red and simulated point sources (MC) in blue.
Shown in the bottom panels are the residuals defined as (MC$-$data) / $\sigma$.
}
\label{IGMFplot}
\end{center}
\end{figure}

The VERITAS observations of seven VHE gamma-ray blazars have allowed for a sensitive search for an angular broadening signature of the IGMF\cite{elisa}.
The best sources for this type of investigation are hard-spectrum, relatively distant point-source blazars that are strongly detected at VHE energies.
The spatial extent of the data for each of these seven objects is compared with the simulated VERITAS point-spread function (PSF).
Figure \ref{IGMFplot} shows this comparison for three of the seven blazars considered here.
No significant evidence of broadening is found.
Model dependent limits are presented for assumed IGMF strengths of B$_{IGMF}$ = $10^{-14}$, $10^{-15}$, and $10^{-16}$ using a three dimensional semi-analytical cascade code\cite{tom}.
The field strengths selected represent a range that VERITAS could potentially be sensitive to, bounded by the limit of $<10^{-16}$, where the expected angular broadening would be too small to detect.
For each field strength and VERITAS object, upper limits are placed on the fraction of the total measured emission arising from cascade emission (as opposed to direct emission).
Based on these ULs, IGMF strengths of $(5-10) \times 10^{-15}$ G are excluded at 95$\%$ confidence level.
Note that a non-detection here does not rule out fields that are strong enough to isotropize the electron/positron pair at the source and only applies to the fields considered here.
Model-independent limits on the flux from extended emission were also calculated assuming a spectral index matching the intrinsic spectral index, resulting 99$\%$ CL upper limits of 0.17$-$2.69$\times$10$^{-12}$ cm$^{-2}$TeV$^{-1}$s$^{-1}$ for energies between 160 GeV and 1 TeV.

Two starburst galaxies (SBGs), galaxies that are characterized by exceptionally high star formation rates (SFRs), have been detected by VHE gamma-ray instruments, M 82 and NGC 253.
These objects provide strong evidence of the correlation between high SFRs and VHE emission, and represent the only known extragalactic point-sources detected at VHE energies without a relativistic jet.
Ultra-luminous infrared galaxies (ULIRGs) are objects predicted to exhibit similar cosmic-ray acceleration and emission mechanisms.
They are objects with high SFRs that contain large amounts of dust, which absorbs the object's UV light, re-radiating it in the FIR.
In \cite{ULIRG}, we show upper limits on VHE gamma-ray emission from two ULIRGs, Arp 220 and IRAS 17208$-$0014.
Our result on Arp 220 represents the most sensitive upper limits on a ULIRG at VHE energies to date and begins to constrain theoretical models.
We further test the correlation between SFR and VHE emission with a deep exposure on the nearby galaxy M 31.
M 31 is the closest spiral galaxy to the Milky Way and contains a dense, star-forming ring which is of particular interest to test this correlation. 
In \cite{M31}, we show a detailed analysis of the detection and morphology of M 31 at lower energies with {\it Fermi}-LAT data and upper limits of the extended object with VERITAS data.

\begin{figure}
\begin{center}
\includegraphics[width = 3.5in]{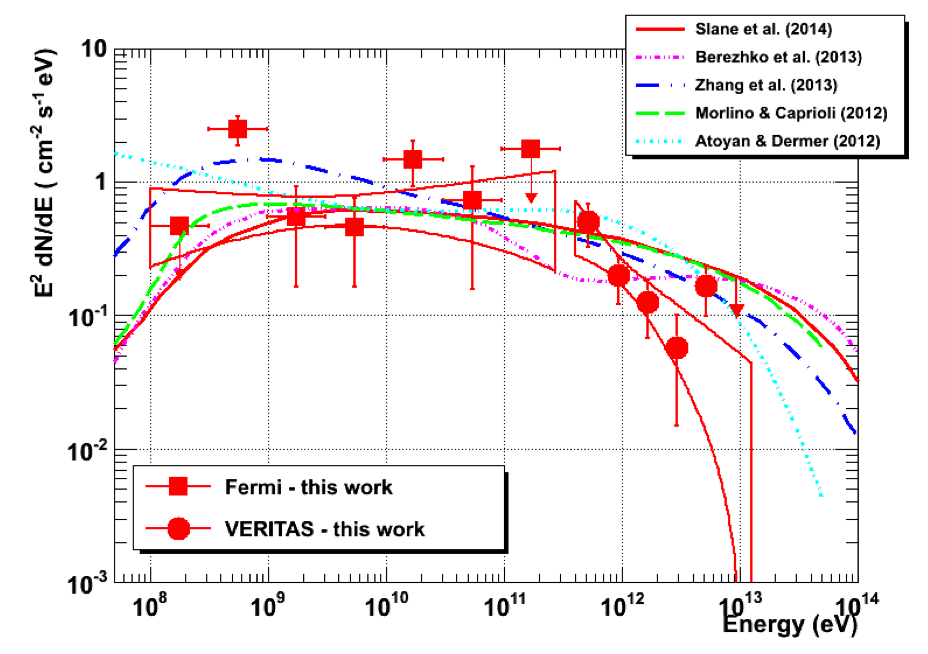}
\caption{
Spectral energy distribution of the Tycho SNR with $Fermi$-LAT and VERITAS data in filled red squares and circles, respectively.
A representative sample of gamma-ray emission models is presented for comparison.
}
\label{tychoPlot}
\end{center}
\end{figure}

\section{Galactic Science}

VERITAS has a rich and varied Galactic program with 20 source detections, including supernova remnants (SNR), pulsar wind nebulae (PWN), high-mass x-ray binaries (HMXB), unidentified sources, and one pulsar.
From its position in the northern hemisphere, VERITAS is able to observe regions of intense star formation like the Cygnus arm, iconic objects like the Crab Nebula and pulsar, the Galactic Center region (above 2 TeV), as well as most of the historical SNRs.
Many of the Galactic objects visible to VERITAS are the most thoroughly studied objects across the spectrum.

At this conference we present several new Crab Nebula and pulsar results.
This includes the first VERITAS measurement on the extension of the VHE emission region of the Crab Nebula, finding no evidence for extension\cite{crabICRC}.
Further, we show contemporaneous VERITAS data taken during the recent dramatic flux variability observed at $Fermi$-LAT energies and demonstrate no variability at VHE energies\cite{crabPaper}.
The Crab pulsar, the power source driving the Nebula energetics, was the first pulsar detected above 100 GeV. 
In these proceedings we present an updated study of the Crab pulsar with twice the data used in the initial VERITAS discovery paper from 2011\cite{crabPulsar}, measuring no cutoff in the energy spectrum to above 400 GeV.
We also show an upper limit on pulsations from the Geminga pulsar from $\sim$72 hours of VERITAS data\cite{geminga} and discuss the status of an archival search of VERITAS data for signals from known $Fermi$ pulsars\cite{fermiPulsar} in data already on disk.

\begin{figure}
\begin{center}
\includegraphics[width = 3.5in]{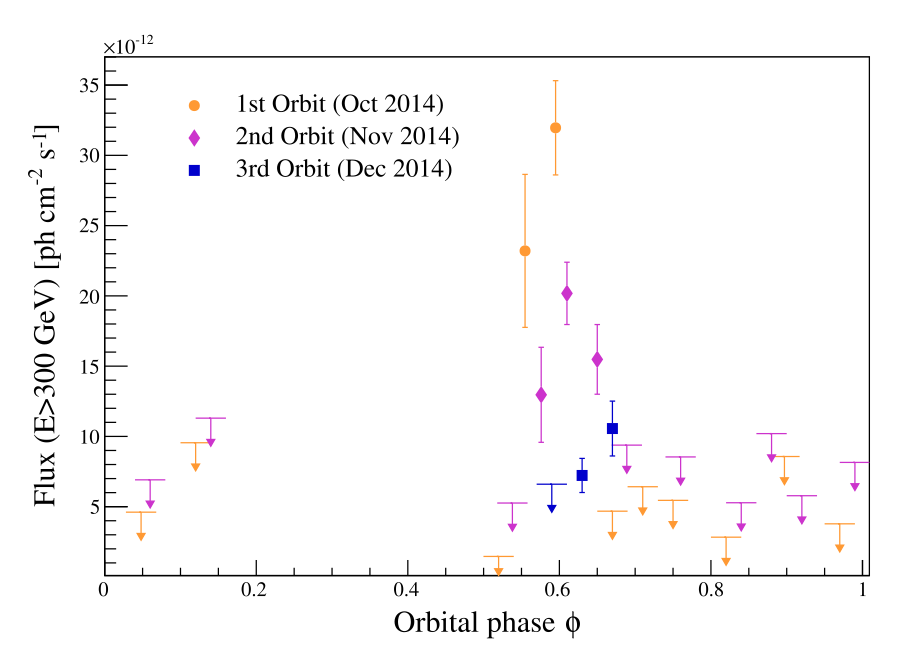}
\caption{
LS I $+$61$^{\circ}$ light curve for three orbits of data taken in the 2014 season.
October is shown as orange circles, November as purple diamonds, and December as blue squares.
Each bin represents a single night with 99$\%$ confidence level flux upper limits shown as arrows.}
\label{lsiPlot}
\end{center}
\end{figure}

Historically, the primary motivation for gamma-ray astronomy was to detect and understand the sites of cosmic-ray acceleration.
The diffusive shocks of supernovae remnants (SNR) remain a plausible and generally accepted explanation.
While X-ray observations have presented clear evidence of the acceleration of electrons to TeV energies, the hadronic acceleration picture is less clear.
A recent $Fermi$-LAT measurement of W 44 and IC 443 at low energies found the expected signature of hadronic emission, evidence of the neutral pion-decay from proton-proton interactions\cite{fermiSNR}.
While hadronic acceleration is favored at VHE energies as well, these energies are less explored. 
To study this VERITAS has committed significant observational time on known SNRs.
In these proceedings, VERITAS has a number of exciting new SNR results based on deep exposures, including presentations on Tycho\cite{tychoP}, Cas A\cite{casA}, and IC 443\cite{ic443}.

Figure \ref{tychoPlot} presents new results for the Tycho SNR, including $\sim$2 times more VERITAS data than previously published.
Tycho provides an excellent laboratory to understand emission mechanisms since it is young, well observed at other wavelengths, and the explosion occurred in a clean environment. 
VERITAS and $Fermi$-LAT data are compared with a selection of favored gamma-ray emission models for Tycho based on the modelling of detailed MWL measurements (most are hadronic-emission dominated).
The camera upgrade of VERITAS led to a lower energy threshold of the instrument and allowed us to measure the Tycho spectrum over a wider range of energies.
The measurement of the updated spectrum is softer than the previously published, $\Gamma = 2.92 \pm 0.42_{stat}$ as opposed to $\Gamma = 1.95 \pm 0.51_{stat} \pm 0.30_{syst}$. 
With this new result and softer spectrum above 500 GeV, data are in tension with the emission scenarios shown.

VERITAS has additionally taken a large exposures on the Cas A and IC 443 SNRs with our upgraded cameras.
In these proceedings, we present data from nearly 60 hours of observations on Cas A, which represents a factor of $\sim$3 more than the published data set.
This data also includes a fraction of the observations taken at large zenith angles (LZAs).  
LZA observations require air showers to traverse a larger atmospheric slant depth, providing an increased detection area by a factor of 5 at high energies (above 1-2 TeV).
No extension is seen for the remnant and the VERITAS spectrum has been extended on both ends, now covering 400 GeV to 7 TeV.
We also present data from 155 hours on IC 443.
VERITAS can now resolve emission structure on the few-arcmin scale and we show new results on the emission morphology of this object, including emission from the shell of the remnant and from other nearby gaseous structures.

Detecting and understanding the VHE gamma-ray emission from high mass x-ray binary (HMXB) systems is another important element of the VERITAS Galactic science program.
These are complex binary systems where VHE emission can be powered by accretion or arise from colliding winds.
There are currently five such systems detected at TeV energies, three of them by VERITAS: LS I +63$^{\circ}$ 303, HESS J0632+057, and LS 5039.
VERITAS commits time to both continued observations on previously detected objects as well as to search for new objects using poor weather observations.
LS I +63$^{\circ}$ 303 is a system that contains a compact object (black hole or neutron star) orbiting a large main sequence star in a $\sim$26.5 elliptical orbit.
This system has been studied extensively across the spectrum and VERITAS has now collected data on it every year since 2006 (totalling over 180 hours).
VERITAS typically detects emission near apastron at the $\sim$10$\%$ Crab Nebula flux level, though there have been years where this source wasn't detected at all.
Conversely, MeV/GeV emission is typically detected throughout the orbit.
VERITAS recently published an energy spectrum for this system with contemporaneous 2011-2012 $Fermi$-LAT data\cite{LSIpaper}.
The spectrum exhibited a cutoff at GeV energies that is typical of the $Fermi$ pulsar population, suggesting separate emission populations at GeV and TeV energies.
We can report here the brightest VHE detection of LS I +63$^{\circ}$ 303 to date\cite{lsiAtel}.
Figure \ref{lsiPlot} shows data from October to December in 2014 sampling three separate orbits.
On October 18th the source was detected with a flux of (31.9 $\pm$ 3.4$_{stat}$) $\times$ 10$^{-12}$ cm$^{-2}$ s$^{-1}$, roughly $30\%$ of the Crab Nebula flux.
The flare rise and fall times are found to be very quick, less than a day, and subsequent orbits were detected closer to the usual flux level indicating some level of orbit-to-orbit variation.
Very fast changes in the flux point to rapid changes in the conditions for acceleration in the system, testing existing models of the emission.
For example, wind-wind collision zones must change quite rapidly.

\begin{wrapfigure}{R}{0.52\textwidth}
\begin{center}
\includegraphics[width = 3.2in]{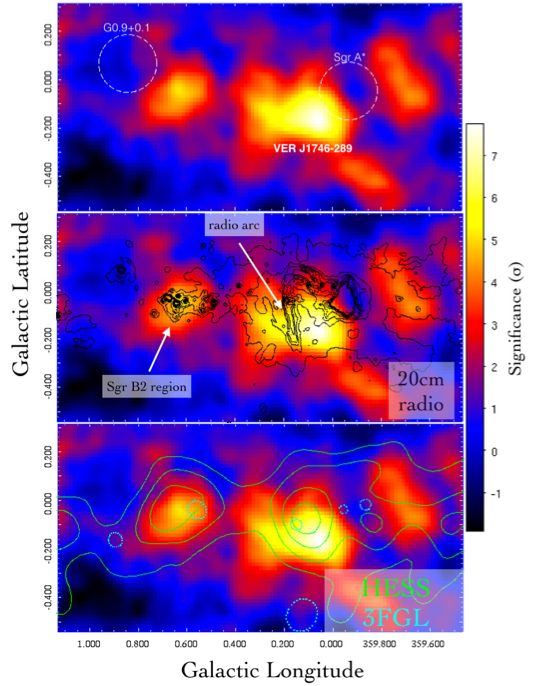}
\caption{Significance maps of the Galactic Center region using VERITAS data with an energy threshold of $\sim$2 TeV.
Shown in the top panel are the locations of the subtracted point sources, Sgr A$^{*}$, and G0 9+0.1, as well as the VERITAS source, VER J1746-289.
Shown in the middle panel is the significance map with overlays of VLA 20cm radio contours.
Shown in the bottom panel is the significance map with overlays of H.E.S.S. excess events and $Fermi$-LAT 3FGL sources.
 }
\label{gcPlot}
\end{center}
\end{wrapfigure}

Despite the fact that the Galactic center (GC) region never rises above 30$^{\circ}$ elevation for VERITAS, we make an exception to observe it.
In fact, as described above, LZA observations give us an effective area boost above 2 TeV compared to standard observations.
We present high statistics measurements of the two known point sources in the GC field, Sgr A$^{*}$, the central supermassive black hole object, and G0 9+0.1, a SNR\cite{andyGC}.
The spectrum of both objects are measured with better statistics at higher energies than prior measurements.
A single power-law fit is found to well describe the G0 9+0.1 spectrum to energies of $\sim$20 TeV, potentially constraining leptonic emission models for this object.
For Sgr A$^{*}$, both a broken power-law and power-law with an exponential cutoff can explain the data over the range of 2$-$30 TeV, but this measurement helps to improve our understanding of the cutoff energy.
After removing both of these point sources from the GC field, a residual sky map shows a rich field of ridge emission $>$2 TeV, as well as evidence for the existence of a new extended source, VER J1746-289.
This ridge emission structure is shown in Fig. \ref{gcPlot}.
It extends west of Sgr A$^{*}$ with several local statistically significant enhancements, including VER J1746-289.
VER J1746-289 is detected at a significance of 7.8$\sigma$ in these maps and is likely correlated with known non-thermal structures in the region.

\section{Dark Matter Science}

The existence of Dark Matter (DM) has been confirmed by observational data at various wavelengths and at various astrophysical scales.
Among the leading candidates for a particle interpretation of DM, Weakly Interacting Massive Particles (WIMPs) present a well-motivated and generic extension to the Standard Model of particle physics.
WIMPs are predicted to produce photon final states from both direct annihilation and/or decay channels as well as through hadronic or leptonic decay chains (photons arise via final state radiation in leptonic chains).
The predicted gamma-ray distribution depends on the M$_{WIMP}$, and for masses above $\sim$100 GeV IACT arrays like VERITAS are sensitive to these signals.
IACT experiments probe regions of theoretical phase space inaccessible to nuclear recoil and accelerator-based experiments.
Additionally, any new particle discovered in a direct detection experiment or at the LHC will need to be further confirmed as the astrophysical DM, an advantage observatories like VERITAS bring to the table.

The VERITAS DM program focuses its search for DM signals in a variety of object classes, each with its own advantages and disadvantages.
The targets include the GC, Galactic DM subhalo candidates, galaxy clusters, and dwarf spheroidal (dSph) galaxies. 
For the purposes of this proceedings, we will focus on the new measurements presented at this conference.

\begin{center}
\begin{figure}
\includegraphics[width = 3in]{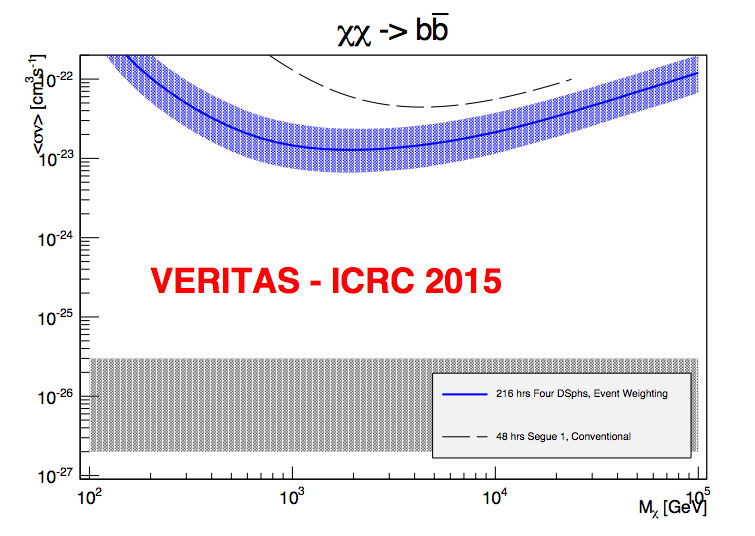}
\includegraphics[width = 3in]{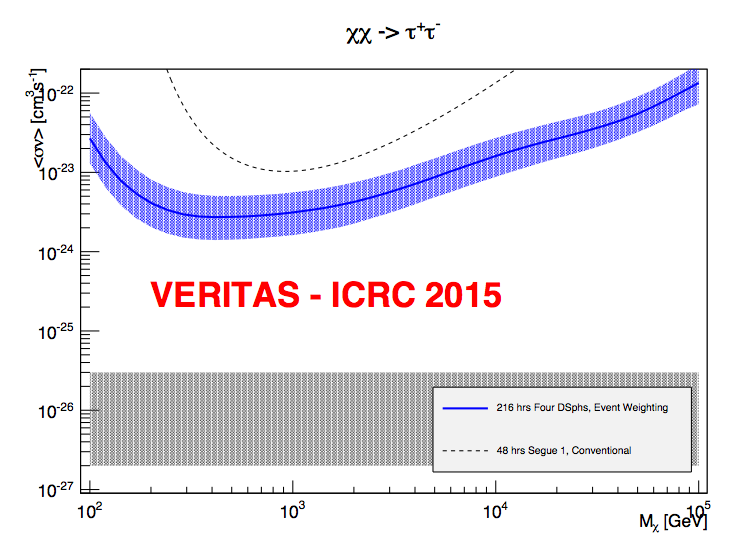}
\caption{
Expected annihilation cross section limits for the joint analysis of dwarf data as a function of DM particle mass.
Limits shown are for 95$\%$ confidence level for two channels, bb (left) and $\tau \tau$ (right), and prior Segue 1 results are also displayed.
The 1$\sigma$ systematical uncertainty on the DM density profile is represented as a blue band and the natural cross section is shown as a grey band for comparison.
}
\label{dmPlot}
\end{figure}
\end{center}

VERITAS currently devotes 150 hours of observation time per year to dSph galaxies.
DSph galaxies are objects of intense study because they are relatively nearby (20-200 kpc), are DM dominated - hosting $\mathcal{O}(10^{3}$) times more DM mass than visible matter, and have no predicted astrophysical gamma-ray signals.
VERITAS has previously published limits on observations of individual dSph sources, the most constraining arising from a 48 hour exposure on Segue I\cite{segue}\cite{erratum}.
However, DM physics is the same across different dSphs, so this topic benefits from an analysis that includes all dSph objects cohesively.
Figure \ref{dmPlot} shows results from a combined analysis using data from five dSph fields observed by VERITAS (Segue 1, Draco, Ursa Minor, Bootes, and Willman 1), comprising $\sim$230 total hours of observation\cite{dmStack}.
These results are an extension of the techniques originally developed for use on $Fermi$-LAT data\cite{alexPaper}.
It is based on an event weighting method that assigns each event a weight based on three factors: the dSph field the event came from, the energy of the event, and the reconstructed position of the event relative to the dSph.
An optimal test statistic is then defined as the sum of all these event weights.
This statistic is used to test a background plus DM hypothesis against the null (background only) for the full dSph dataset.
Figure \ref{dmPlot} presents the exclusion limits from this analysis for two annihilation channels.
This represents the first joint analysis using all VERITAS dSph data at once and improves our dSph DM constraints over previous measurements significantly.
In this work, we also present VERITAS ULs on a search for direct annihilation gamma-ray lines in the dSph data.

High resolution simulations of a DM halo similar to the Milky Way have found DM substructures clumped at all scales.
Not all of these subhalos are predicted to have accumulated enough baryonic matter to be visible, and some in fact may be detectable at gamma-ray energies alone.
VERITAS has enacted a program to select a small sample of the best motivated DM subhalo candidates from $Fermi$-LAT unassociated objects to follow-up with VHE energies.
Candidates were selected based on multi-wavelength properties, lack of time variability, and detectability with VERITAS.
In \cite{nieto} we present VERITAS upper limits from the observations of two of these objects, 2FGL J0545.6+6018 and 2FGL J1115.0$-$0701.
We also discuss possible DM model scenarios utilizing the VERITAS dataset and 7 years of $Fermi$-LAT data.

\section{Cosmic-ray Measurements}

While VERITAS is primarily a gamma-ray instrument, VERITAS data can also be used to make several high-impact cosmic-ray measurements.
VERITAS benefits from the fact that cosmic-ray electron (CRE) and cosmic-ray iron (CRI) events are diffuse and isotropic signals on the sky and are therefore buried within the background of all observed fields.
Due to the diffuse nature of these signals, the standard analysis methods that use defined ON and OFF regions within the same field of view do not work.
In these proceedings, we present a CRE spectrum with VERITAS data \cite{CRE}.
We also discuss the status of a VERITAS program to measure the separated electron and positron fluxes (as well as the proton and anti-proton fluxes) using the imprint of the moon shadow on the isotropic cosmic-ray flux\cite{moonShadow} and the progress towards a new measurement of the CRI spectrum above 10 TeV\cite{CRH}.

\begin{figure}
\begin{center}
\includegraphics[width = 3.5in]{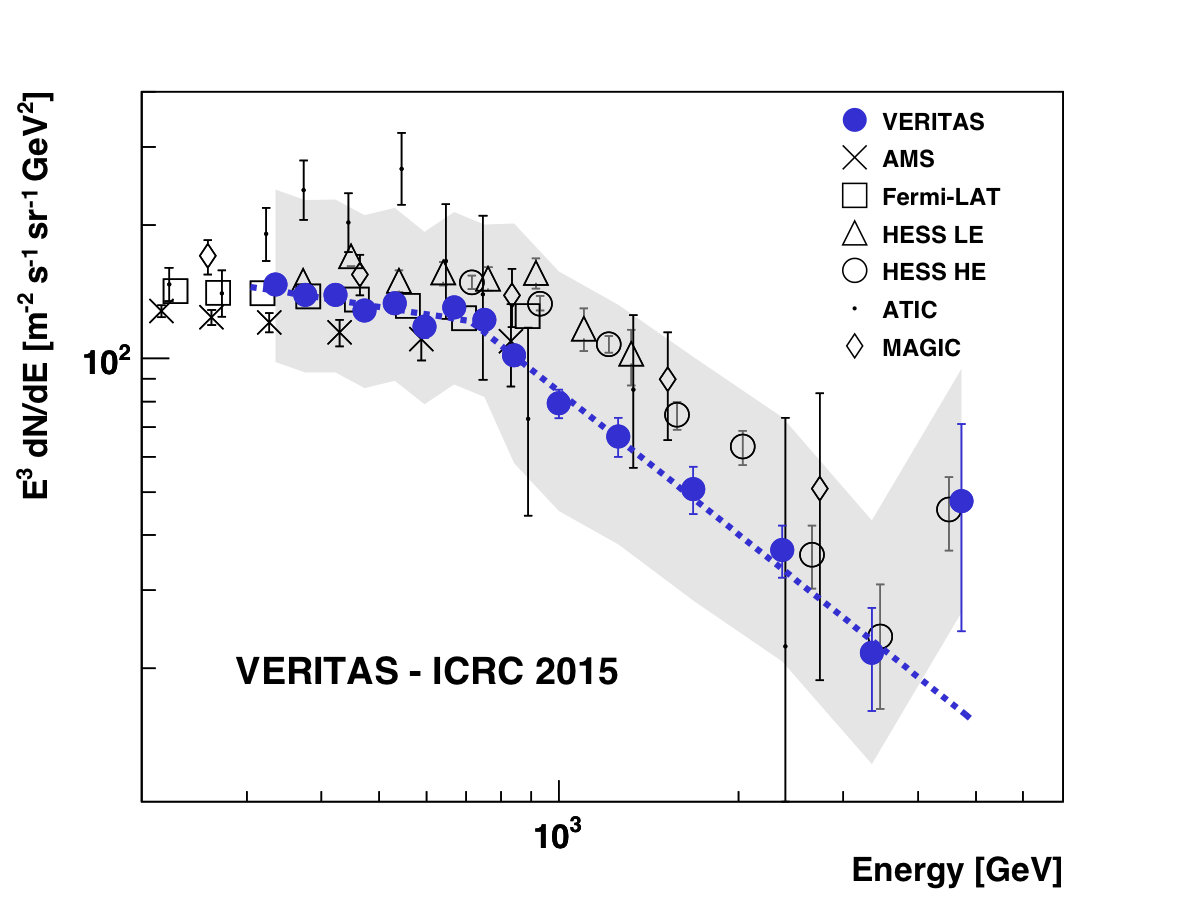}
\caption{
Preliminary cosmic-ray electron spectrum from $\sim$290 hours of VERITAS data covering the $\sim$300 GeV to $\sim$5 TeV energy range.
Also shown are other satellite-based and ground-based measurements with overlapping energy data points.
The best fit to the VERITAS data is shown as an overlaid dashed line.
The gray band represents the systematical uncertainty.
}
\label{crePlot}
\end{center}
\end{figure}

To isolate CRE events in our data and form an energy spectrum, we first use a boosted decision tree discriminator to quantify events as more signal-like or background-like.
The electron fraction of the data is then extracted using an extended likelihood fitting technique.
In Fig. \ref{crePlot}, we show the preliminary VERITAS CRE energy spectrum with VERITAS data\cite{CRE} as compared with prior measurements that pioneered this technique using IACT data\cite{hessCRE}.
The result qualitatively agrees with prior satellite-based and ground-based measurements of the CRE spectrum within systematical uncertainties.  
The best fit for the data is a broken power-law with spectral index of $-$3.2 $\pm$ 0.1$_{stat}$ ($-$4.0 $\pm$ 0.1$_{stat}$) below (above) a measured cutoff energy of 710 $\pm$ 40 GeV.
Of the many satellite-based and ground-based experiments studying CREs, this measurement represents only the second high statistics measurement a cutoff at an energy of a $\sim$TeV.

\section{Gamma-ray Bursts}

In the VERITAS observational program, gamma-ray bursts (GRBs) are given the highest priority among all sources if they are visible by the array within $\sim$3 hours of their detection.
GRB 130427A was an unusually bright and nearby (z=0.34) gamma-ray burst detected by many satellite- and ground-based observatories.
In particular, the $Fermi$-LAT detected gamma rays from the burst for up to $\sim$70 ks after the initial detection.
VERITAS began observations of this burst $\sim$71 ks (about 20 hours) after the initial detection but did not detect any significant emission $>$100 GeV\cite{grbPaper}.
It is unfortunate that the initial night of the burst was during the full moon and VERITAS was not observing because an extrapolation of the $Fermi$-LAT spectrum into VHE predicts $\sim$100 $\gamma$/s during the prompt phase of the GRB.
Due to the close proximity of this burst and the favorable zenith observational conditions, VERITAS observations 20 hours after the burst still constrain several emission models.
Both synchrotron and inverse Compton scenarios that attempt to explain the late-time, high-energy emission detected by the LAT are constrained by the VERITAS ULs.

We also present a new approach to the analysis of GRB data in these proceedings\cite{grbPro}.
This approach improves upon the standard Li and Ma approach by statistically accounting for the $a$ $priori$ knowledge of the time dependence of the source light curve.
Based on a Monte Carlo study, we determined that this method improves our sensitivity to GRB-like signals and has been used to reanalyze a sub-sample of VERITAS GRBs.

\section{IceCube Events}
	
The IceCube discovery of an astrophysical flux of high energy neutrinos provides evidence of the interaction of energetic hadrons and points to the sites of the acceleration of high energy cosmic rays\cite{IceCube}.
However, IceCube events to date appear isotropic and no neutrino point-sources have been discovered.
For the past two years VERITAS has been observing the positions of IceCube events, focusing on $\nu_{\mu}$-induced muon track events (events that have the best angular uncertainties, $\sim$1$^{\circ}$).
The list of observed positions includes 3 published events and 19 unpublished events that have been shared with VERITAS through a cooperation agreement.
Observations of these positions show no statistically significant gamma-ray signals and we calculate upper limits in the range of a few percent of the Crab Nebula flux above 100 GeV for each position\cite{IC}.

\section{Conclusion and Outlook}

VERITAS has recently completed its eighth successful year of operations, and the scientific output of VERITAS since the last ICRC conference remains strong.
This is made possible by many factors, including the dedicated operations work of the local FLWO staff, a shift in the focus of the VERITAS observing program from source discovery to deep exposures on scientifically interesting objects, as well as the cultivation of several strong MWL partnerships.
Many of the outstanding issues in VHE astrophysics will only be solved with MWL data.
Several long-term MWL campaigns are underway to contemporaneously study the nature of AGN jets using instruments across the spectrum.
In particular, we benefit from simultaneous observations by X-ray instruments, like $Swift$ and $NuSTAR$, since they are sensitive to the same non-thermal, relativistic populations of particles that produce gamma-ray emission.
The Mrk 501 MWL campaign presented at this conference represents the most complete broadband campaign on this object to date, with several absolutely simultaneous snapshots allowing SED reconstruction from radio through VHE gamma-ray energies\cite{501}.
Additionally, HAWC and VERITAS are particularly well matched for collaboration since they view essentially the same region of the sky at the same time.
The HAWC experiment recently completed their full array, and annual meetings have been set up between HAWC, VERITAS, and $Fermi$-LAT to increase communication and collaboration between the experiments.

In addition to upgrading hardware, VERITAS is implementing new analysis techniques to improve the sensitivity of our instrument.
In these proceedings we show several results using advanced analysis methods as a replacement of the standard Hillas box cuts, including the use of boosted decision tree multivariate analyses\cite{nahee}\cite{elisa}\cite{CRE}, a template method\cite{template}\cite{CRH}, and HFit, a 2D gaussian image fitting technique\cite{dmStack}. 
These first results represent an avenue that VERITAS will continue to improve, building techniques that will become standard in VERITAS analyses.

Finally, the results presented here are only a subsample of the science coming from VERITAS for the last two years.
VERITAS plans to operate at least until 2019, giving us the opportunity to utilize overlap with new instruments like HAWC and {\it NuSTAR} and to continue to operate until the next generation of ground-based arrays come online.

\section*{Acknowledgements}

This research is supported by grants from the U.S. Department of Energy Office of Science, the U.S. National Science Foundation and the Smithsonian Institution, and by NSERC in Canada. We acknowledge the excellent work of the technical support staff at the Fred Lawrence Whipple Observatory and at the collaborating institutions in the construction and operation of the instrument.

The VERITAS Collaboration is grateful to Trevor Weekes for his seminal contributions and leadership in the field of VHE gamma-ray astrophysics, which made this study possible.

\end{document}